\documentclass[12pt,aps,nofootinbib, superscriptaddress,amsfonts,amsmath,amssymb]{revtex4}

\usepackage{graphicx,amsmath,amssymb}
\usepackage{bm}

\begin{document}

\title{
Analytical approach to relaxation  dynamics of condensed Bose gases
}
\author{Miguel Escobedo}
\author{Federica Pezzotti}
\affiliation
      {%
       Departamento de Matem\'aticas, 
       Universidad del Pa\'\i s Vasco, 
       Apartado 644, E-48080 Bilbao, Spain
       }%
\author{Manuel Valle}
\affiliation
      {%
       Departamento de F\'\i sica Te\'orica, 
       Universidad del Pa\'\i s Vasco, 
       Apartado 644, E-48080 Bilbao, Spain
       }%

\date{\today}

\begin{abstract}
The temporal evolution of a perturbation of the equilibrium distribution of a condensed Bose gas is
investigated using the kinetic equation which describes collision between condensate 
and noncondensate atoms. The dynamics is studied in the low momentum limit 
where an analytical treatment is feasible.  
Explicit results are given for the behavior at large times in different temperature regimes. 
\end{abstract}



\maketitle


\section{\label{sec:intro} Introduction}

A distinctive feature of the quantum kinetic theory of the condensed Bose gas 
arises from the correlations between the superfluid  component and the normal fluid part  
corresponding to the excitations. This causes 
the occurrence of number-changing processes determining 
the relaxation to equilibrium when excitations collide frequently. 
As a consequence, in the hydrodynamic regime,  
a collision integral $C_{12}$ describing $1 \leftrightarrow 2$ splitting 
of an excitation into two others 
in the presence of the condensate is needed. 
The specific form of such a term depends on the dispersion relation $\omega(k)$
for the energy of quasiparticles and 
on the matrix element $\mathcal {M}$ of the effective 
Hamiltonian describing the interaction between them. 

In a series of papers before the experimental realization of BEC, 
Kirkpatrick and Dorfman~\cite{Kirkpatrick1,Kirkpatrick2}
have derived a kinetic equation in a uniform Bose gas 
which includes these processes, 
and they have computed explicit values for the transport 
coefficients of the two-fluid hydrodynamics.  
More recently, Zaremba, Nikuni and Griffin~\cite{Zaremba}
have extended the treatment to a trapped Bose gas 
by including  
Hartree-Fock corrections to 
the energy of the excitations, and they 
have derived  coupled kinetic equations for the distribution
 functions of the normal  and superfluid components. 
Many implications of their approach concerning 
transport coefficients and relaxation times 
have been thoroughly reported in Ref.~\cite{book2}, 
mainly when $k_\mathrm{B}T \gg g n_\mathrm{c}$, where $g = 4 \pi a m^{-1}$  
is the interaction coupling constant in 
terms of the $s$-wave scattering length $a$, 
and $n_\mathrm{c}$  is the condensate density. 
In this regime the
 dispersion law approaches the free particle law 
$\omega(k) \approx k^2/(2 m) $. 
In the opposite 
low temperature limit $k_\mathrm{B}T \ll g n$,  the relevant  
part of the excitation spectrum  is $ \omega(k) \approx c k$, 
where $c = \sqrt{g n m^{-1}}$  is the sound speed 
near zero temperature and $n$ is the particle density.  
The kinetic equation in this regime has been derived  
in  Ref.~\cite{Imamovic},  
but its features  have been  less considered. 
Due to  the absence of studies on the eigenvalue problem posed by 
the linearized kinetic equation, 
little is known  about the time behavior of the solutions of  
kinetic equations in both regimes. 

This paper is devoted to deriving some  
results on the free evolution of initial perturbations  of thermal equilibrium as 
described by the three-excitation collision term near the 
critical and near zero temperature. 
By making  appropriate approximations of 
the kinetic equation we shall see that the 
problem becomes analytically  tractable in some limits, 
and hence the asymptotic behavior of the solutions for large time 
may be evaluated.
In the first regime, $k_\mathrm{B}T \gg g n_\mathrm{c}$,  
we consider 
the small momentum limit where the equilibrium distribution function 
obeys the condition $n_0(k) \gg 1$. 
By approximating the full linearized kinetic equation in this limit 
we obtain an integro-differential equation which, after integration by parts,  
agrees with 
linearized evolution equation of 
wave turbulence~\cite{Newell,book} with the appropriate indices. 
This result is not a priori obvious when one starts from 
the full kinetic equation, 
because of the strong singularity of the integrand.  
At low temperature,  $k_\mathrm{B}T \ll g n$, 
we consider two opposite regimes $n_0(k) \gg 1$ and $n_0(k) \ll 1$. 
By performing a low momentum approximation, $c k \ll k_\mathrm{B}T$, 
we will show that the wave turbulence framework still emerges 
but in a form not consistent with energy conservation, so 
an additional improvement is needed in order to restore this  conservation law. 
In the opposite regime near zero temperature, 
where $c k \gg k_\mathrm{B}T$,  we shall consider 
an approximation based on the dominance of Beliaev damping processes.

A notable feature  of the linearized collision terms 
in the wave limit 
is the homogeneous dependence on the momentum.    
Such a property furnishes a systematic way to 
compare  the relative orders  of 
different collision integrals at low momentum. 
This is achieved 
by  simply considering the degrees  of homogeneity,  since their values  
depend on some indices related to 
the dispersion law, the scattering amplitude and the number of bodies in the collision. 
Thus one can see the dominance  
of $C_{12}$ over the binary collision term $C_{22}$.

The plan of the paper is the following. In Sec.\ref{sec:lin} we discuss the general form  
of the linearized kinetic equation as written in terms of irreducible components 
of the perturbation. Sec.\ref{sec:crit} is first devoted to deriving 
the low momentum approximation of the evolution equation below the critical temperature.
Then, once we show that the wave turbulence picture is recovered, 
we briefly review the general procedure for solving this kind of equation, 
we apply these techniques to the present context 
and we determine the asymptotic behavior of the perturbation for large time.
Anisotropic perturbations are also discussed. 
In Sec.~\ref{sec:class} we derive the evolution equation in the thermal regime,  
with particular attention to the requirement  of energy conservation, and 
we deal with the issue of  the dominance of $C_{12}$ over $C_{22}$. 
In Sec.~\ref{sec:qua} we consider the case of very low temperature and
derive some asymptotic results specific for this regime. 
Section~\ref{sec:end} contains some concluding remarks.


\section{\label{sec:lin} Linearized collision integral in the Fredholm form} 

The form of the three-excitation collision integral 
\begin{equation}
\label{eq:kin}
C_{12}[n]=  \int \left[ R(\bm{k}, \bm{k}_1, \bm{k}_2) -
R(\bm{k}_1, \bm{k} , \bm{k}_2)-R(\bm{k}_2, \bm{k}_1, \bm{k})\right]  d^3 \bm{k}_1  d^3 \bm{k}_2, 
\end{equation}
depends essentially on
the dispersion law $\omega(k)$ 
and the matrix element $\mathcal{M}$ of the three-excitation interaction. 
These quantities determine $R$ by means of 
\begin{eqnarray}
R(\bm{k}, \bm{k}_1, \bm{k}_2) &=&  |\mathcal M(\bm{k}, \bm{k}_1, \bm{k}_2)|^2 \left[
\delta(\omega(\bm{k}) - \omega(\bm{k}_1) - \omega(\bm{k}_2))
\delta(\bm{k} -\bm{k}_1 - \bm{k}_2)\right] \nonumber \\ 
&& \times \left[ n_1 n_2 (1+n) - (1+n_1) (1+n_2) n \right] . 
\end{eqnarray}
In this work we will consider two extreme regimes of the Bogoliubov dispersion law
\begin{equation}
 \omega(k) = \left[\frac{g n}{m}  k^2  + \left(\frac{k^2}{2m}\right)^2 \right]^{1/2} , 
\end{equation}
near the critical temperature and near zero temperature. 
In the first limit  $k_\mathrm{B}T \gg g n_\mathrm{c}$ the dispersion law and the 
squared amplitude become~\cite{Zaremba} 
\begin{eqnarray}
\label{eq:crit1}
\omega(k) &\approx & \frac{k^2}{2 m}+ g n_\mathrm{c}, \\ 
\label{eq:crit2}
|\mathcal M(\bm{k}, \bm{k}_1, \bm{k}_2)|^2 &=& 
\frac{ g^2 n_\mathrm{c}}{2 \pi^2} =\frac{8 n_\mathrm{c} a^2}{m^2}  , 
\end{eqnarray}
where $n_\mathrm{c}$ is the superfluid density.
Since in this regime the average value of the energy of a quasiparticle is  
of order $k_\mathrm{B}T$ 
the term $g n_\mathrm{c}$ is subleading in Eq.~(\ref{eq:crit1}).
In the opposite limit $k_\mathrm{B}T \ll g n$  we have~\cite{Imamovic} 
\begin{eqnarray}
\label{eq:low1}
\omega(k) &\approx & c  k + \alpha k^3, \\ 
\label{eq:low2}
|\mathcal M(\bm{k}, \bm{k}_1, \bm{k}_2)|^2 &=& \frac{9 g k k_1 k_2}{64 \pi^2 m^2 c}  = 
 \frac{9 c k k_1 k_2}{64 \pi^2 m n} . 
\end{eqnarray}
The amplitudes $\mathcal{M}$ can be derived heuristically 
by golden rule arguments from  the effective Hamiltonians in 
these regimes~\cite{Eckern}.
Note that no distinction is made between $n$ and $n_\mathrm{c}$  in 
the scattering amplitude valid at $T \rightarrow 0$,  
because
we neglect the small low temperature depletion. 
Since we are interested in the low momentum region,  
we shall retain the first term of the dispersion law. 
Here we will consider  the homogeneous regime where 
the distribution function is independent of the position. 

The linearization of  the  kinetic equation $\partial_t n = C_{12}[n]$ 
proceeds by the insertion of 
\begin{equation}
 n(\bm{k}, t)  =  n_0(\omega(k))  + n_0(\omega(k))[1 + n_0(\omega(k))] \chi(\bm{k}, t) , 
\end{equation}
where $n_0(\omega(k))$ is the Bose distribution function  without chemical potential with 
the energy given by the first term of 
Eqs.~(\ref{eq:crit1}) or (\ref{eq:low1});  
the dimensionless function $\chi(\bm{k}, t)$ parametrizes the departure from equilibrium for 
the noncondensate distribution function. 
The linearized kinetic equation adopts the form
\begin{equation}
n_0(1 + n_0) \frac{\partial \chi}{\partial t} = L[\chi](\bm{k}, t)  = 
 \int \mathcal{L}(\bm{k}, \bm{k}') \chi(\bm{k}', t) d^3 \bm{k}'  ,  
\end{equation}
which must be completed with 
the linearized equation for the  fluctuation of the condensate  density 
$\delta n_\mathrm{c}$,
\begin{equation}
\label{eq:nonc}
\frac{d \delta n_\mathrm{c}}{d t} = -\delta \Gamma_{12} = -\int   L[\chi](\bm{k}) \frac{d^3 \bm{k}}{(2 \pi)^3} . 
\end{equation}
The  kernel $\mathcal{L}$ is symmetric  in $\bm{k}$ and $\bm{k}'$ and, 
due to the Bose functions, falls off sufficiently  as $k, k' \rightarrow \infty$.  
We may then introduce the following scalar product
\begin{equation}
(\chi_1, \chi_2)  =  \int n_0(1+n_0) \chi_1(\bm{k}) \chi_2(\bm{k}) d^3 \bm{k} , 
\end{equation}
and pose the eigenvalue problem
\begin{equation}
 \int \mathcal{L}(\bm{k}, \bm{k}') \chi_j(\bm{k}') d^3 \bm{k}'   = -\lambda_j n_0(1+n_0) \chi_j(\bm{k}) . 
\end{equation}
The solution of the linearized equation  would be formally written as the following series 
\begin{equation}
\chi(\bm{k}, t) = \sum_j \ c_j \chi_j(\bm{k}) e^{-\lambda_j t} , 
\end{equation}
where the coefficients $c_j$ would be determined by the initial condition. 
In this work we do not  
intend to follow this procedure since 
it is uncertain that there exists a spectrum of discrete  eigenvalues for this problem. 
This doubt is based on the observation that 
the lifetime of a long-lived well-developed sound mode of energy $\omega = c k$ 
varies continuously with the momentum, $\tau(k) \propto k^{-\delta}$,  
where $\delta=1$ at $T\neq 0$~\cite{Hohenberg}  or  $\delta=5$ at $T=0$~\cite{Beliaev}.
So,  this property does not seem to be consistent with the discrete 
character of the spectrum. 
Instead, we will try to find an expression for 
the action of the operator $L[\chi]$ at low momentum, 
and then solve the evolution equation found in this approximation. 

Formally, the operator $L$ is decomposed in
 two types of contributions
\begin{equation}
L[\chi] = -\mathcal{N}(k) \chi(\bm{k},t) +  \int \mathcal{U}(\bm{k}, \bm{k}') \chi(\bm{k}', t) d^3 \bm{k}' , 
\end{equation}
where the function $\mathcal{U}$ is given by 
\begin{eqnarray}
\label{eq:UU}
\mathcal{U}(\bm{k}, \bm{k}') &=& \left[
 2|\mathcal{M}(\bm{k}, \bm{k}', \bm{k}-\bm{k}')|^2 
  \delta(\omega(\bm{k}) - \omega(\bm{k}' )- \omega(\bm{k}-\bm{k}')) \right. \nonumber  \\ 
 && \times \left. 
n(\omega) [1+n(\omega')] [1+n(\omega-\omega')]  + (\bm{k} \leftrightarrow \bm{k}') \right]  \nonumber  \\  
 &&-2|\mathcal{M}(\bm{k}+\bm{k}', \bm{k}, \bm{k}')|^2  
 \delta(\omega(\bm{k}) + \omega(\bm{k}' )- \omega(\bm{k}+\bm{k}'))\nonumber  \\ 
 && \times 
[1+n(\omega)] [1+n(\omega')] n(\omega+\omega') , 
\end{eqnarray}
and the explicit form of the coefficient  $\mathcal{N}(k)$ will be not required. 

The conservation laws of energy and momentum imply 
the presence of zero modes of $L$ proportional to the energy and 
the momentum, $\chi(\bm{k}, t) \propto  \omega(k),  \bm{h}(t) \cdot \bm{k}$. 
In particular, 
\begin{equation}
L[\beta \omega(k)]= -\mathcal{N}(k) \beta \omega(k) +  
\int \mathcal{U}(\bm{k}, \bm{k}') \beta \omega(k') d^3 \bm{k}' = 0. 
\end{equation}
Thus, the linearized collision operator can be written as 
\begin{equation}
L[\chi]  = \int \mathcal{U}(\bm{k}, \bm{k}') \left[ \chi(\bm{k}', t) -
\beta \omega(k') \frac{\chi(\bm{k}, t)}{\beta \omega(k)} \right] d^3 \bm{k}' , 
\end{equation}
which suggests to introduce a dimensionless variable $A(\bm{k},t)$ defined as 
\begin{equation}
A(\bm{k},t) =  \frac{\chi(\bm{k}, t)}{\beta \omega(k)} . 
\end{equation}
In terms of $A$,  the linearized kinetic equation has the form
\begin{equation}
n_0 (1+n_0) \beta \omega(k) \partial_t A(k, t)  =  
\int \mathcal{U}(\bm{k}, \bm{k}')  \beta \omega(k') \left[ A(\bm{k}', t) -A(\bm{k}, t) \right] d^3 \bm{k}'  . 
\end{equation}
The rotational invariance of the kernel $\mathcal{U}(\bm{k}, \bm{k}')$  
may be exploited by  
expressing the  perturbation as a superposition of angular  momentum eigenstates
\begin{equation}
A(\bm{k}, t)  =  \sum_{l, m} \mathcal{A}_{l m}(k,t) Y_{l m}(\hat{\bm{k}}) .
\end{equation}
Using the addition theorem for spherical harmonics one arrives at the equation
\begin{equation}
\label{eq:kin2}
n_0 (1+n_0) \beta \omega(k)\partial_t \mathcal{A}_{l m}(k, t) = K_l[\mathcal{A}_{l m}](k, t)
\end{equation}
where the operator $K_l$ adopts the Fredholm form 
\begin{eqnarray}
\label{eq:fred}
K_l[\mathcal{A}] (k, t)&=& \frac{4 \pi}{2 l+1} \int_0^\infty \mathcal{U}_{l}(k, k') 
\beta \omega(k')
 \left[\mathcal{A}(k', t) -  \mathcal{A} (k, t) \right] k'^2 dk' \nonumber \\
 && -  4\pi  \mathcal{A}(k, t)\int_0^\infty 
 \left( \mathcal{U}_{0}(k, k') -  \frac{\mathcal{U}_{l}(k, k')}{2 l+1}  \right)\beta \omega(k') 
k'^2 dk' . 
\end{eqnarray}
Here $\mathcal{U}_{l}(k, k')$  is the $l$-coefficient in 
the series  of Legendre polynomials of $\mathcal{U}$: 
\begin{equation}
\mathcal{U}(\bm{k}, \bm{k}') =  \sum_{l=0}^\infty \mathcal{U}_{l}(k, k') P_l(\cos \theta_{\bm{k} \bm{k}'}) . 
\end{equation}

Next we derive two approximations to $K_l[\mathcal{A}] (k, t)$ 
at low momentum valid near the critical temperature and low temperature. 
They must formally satisfy energy conservation which is achieved if 
\begin{equation}
\frac{d}{dt} \int_0^\infty n_0 (1+n_0) \beta \omega(k)\mathcal{A}_{00}(k ,t) \omega(k) k^2 dk = 
\int_0^\infty K_0[\mathcal{A}_{00}] (k, t) \omega(k) k^2 dk = 0. 
\end{equation}
Clearly, in view of Eq.~(\ref{eq:fred}), 
the symmetry of the kernel $\mathcal{U}_0(k,k')$ is sufficient in order to accomplish 
this requirement,  so 
such a property  must not be lost in the approximated kernel.

\section{\label{sec:crit}  Dynamics near  the critical temperature} 

\subsection{The linearized equation at low momentum}
\begin{figure}
\centering
    \includegraphics{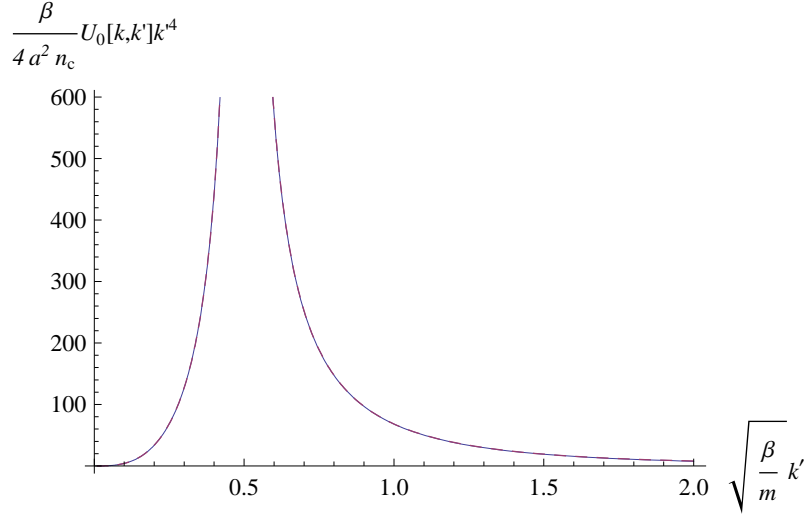}
    \caption{\label{fig:U0T} 
    The kernel $U_0(k, k') k'^4$ for $k\sqrt{\beta/m} =1/2$ 
    near the critical temperature. For this value 
    there is no appreciable difference between the graphs  of the exact kernel and 
    the  approximation of  Eq.~(\ref{eq:U0crit}).
    The singularity at $k'=k$ is not integrable.} 
\end{figure}

In this section we consider the regime $k_\mathrm{B} T \gg g n_\mathrm{c}$ 
near the critical temperature. Here the dispersion is well approximated by 
$\omega = k^2/(2 m)$,  and the squared amplitude by  $|\mathcal{M}|^2 = 8 n_\mathrm{c} a^2 m^{-2}$.
The low momentum approximation to $\mathcal{U}_0(k,k') $ is obtained 
by the rescaling $k \to \epsilon k, k' \to \epsilon k'$ and the expansion around $\epsilon=0$. 
This gives
\begin{equation}
\label{eq:U0crit}
\mathcal{U}_0(k,k') \sim \frac{128 m^2 a^2 n_\mathrm{c}}{\beta^3} 
\left(\frac{\theta(k-k')}{k^3 k' (k^4-k'^4)} + \frac{\theta(k'-k)}{k\, k'^3 (k'^4-k^4)}  \right), \quad k \to 0, \; k' \to 0. 
\end{equation} 
The kernel  $\mathcal{U}_l(k,k' )$ has a non-integrable singularity  proportional to 
$|k-k'|^{-1}$ as $k'  \rightarrow k$ (see Fig.~\ref{fig:U0T}). 
In order to manage carefully this singularity, it is convenient to integrate by parts 
in the exact expression of $K_l[\mathcal{A}] $  given by Eq.~(\ref{eq:fred}). 
It helps to write
\begin{equation}
\mathcal{A}(k', t) -  \mathcal{A} (k, t) =  \int_k^{k'} dq \, \partial_q \mathcal{A}(q,t) ,
\end{equation}
and to perform the change of the order of integration. 
This yields
\begin{eqnarray}
\label{eq:dif}
K_l[\mathcal{A}] (k, t)&=&  \int_0^k dq  \,  J_l^< (k, q) \partial_q \mathcal{A}(q,t)+
\int_k^\infty dq  \,  J_l^> (k, q) \partial_q \mathcal{A}(q,t)
 \nonumber \\
 && -  4\pi  \mathcal{A}(k, t)\int_0^\infty \left( \mathcal{U}_{0}(k, k') -  
  \frac{ \mathcal{U}_{l}(k, k')}{2 l+1} \right)\beta \omega(k')
k'^2 dk'  , 
\end{eqnarray}
where 
\begin{eqnarray}
\label{eq:J1}
J_l^< (k, q) &= & -\theta(k-q) \frac{4 \pi}{2 l+1} \int_0^q \mathcal{U}_l(k,k') \beta \omega(k')  k'^2 dk' ,  \\
\label{eq:J2}
J_l^> (k, q) &= & \theta(q-k) \frac{4 \pi}{2 l+1} \int_q^\infty \mathcal{U}_l(k,k') \beta \omega(k')  k'^2 dk' ,  
\end{eqnarray}
and $\theta(x)$  is the unit step function. 

The behavior of the first term in the right hand side of Eq.~(\ref{eq:dif}) 
as $k \rightarrow 0$ is  derived by 
determining the leading behavior of $J_l^{<}(k,q)$  
as both arguments $k, q \rightarrow 0$.
With respect to the second term in Eq.~(\ref{eq:dif}) 
it is far from obvious that the limit $q \rightarrow 0$ of  $J_l^{>}(k,q)$ must be incorporated 
since $q$  runs  from $k$ to $\infty$. 
The key idea emerges by noting first that the considered scattering amplitude 
involves the assumption that all three momenta are of the same order.
It is then expected to be adequate to take the limit  
$q\rightarrow 0$ when $k \rightarrow 0$ in $J_l^{>}(k,q)$. 
Another more formal justification arises from 
the fact that $J_l^{>}(k,q)$ diverges as $\ln (q-k)$ as $q \to k$. 
Therefore, 
by assuming that 
$\partial_q\mathcal{A}(q,t)$ falls off sufficiently to insure integrability, 
the main contribution to the second term in Eq.~(\ref{eq:dif}) as $k \rightarrow 0$
comes from the region where $q$ is also close to zero.
The  leading behavior of the integral in  the last term of Eq.~(\ref{eq:dif}) as $k \to 0$ 
is easily determined since the singularities of $(2l+1)\mathcal{U}_0$ and $\mathcal{U}_l$ cancel 
each other. 

The evaluation of the integrals in that limit can be carried out analytically,  
and they assume the scaling  form
\begin{eqnarray}
J_l^{<}(k,q) &\sim& \frac{k^{-2}}{q}  \mathcal{H}_l^{<} \left( \frac{k}{q}\right) \theta(k-q), \quad k, q \rightarrow 0 ,  \\ 
J_l^{>}(k,q) &\sim& \frac{k^{-2}}{q}  \mathcal{H}_l^{>} \left( \frac{k}{q}\right) \theta(q-k), \quad k, q \rightarrow 0 .
\end{eqnarray}
Explicit expressions of $\mathcal{H}_l^{<, >}(x)$ will be given below. 
In this approximation the action of the operator $K_l$ is written as   
\begin{eqnarray}
\label{eq:asymp0}
K_l[\mathcal{A}] (k, t) &\sim&  k^{-2}  
\int_0^\infty \mathcal{H}_l\left( \frac{k}{q}\right) \partial_q \mathcal{A}(q,t) \frac{dq}{q}
 - k^{-3} \mathcal{I}_l \mathcal{A}(k,t) , \quad k \rightarrow 0,
 \end{eqnarray}
 where 
 \begin{equation}
 \mathcal{H}(x) =  \mathcal{H}_l^{>}(x) \theta(1-x) + \mathcal{H}_l^{<}(x) \theta(x-1). 
 \end{equation}
Note that $\mathcal{I}_0 =0$. 
Since the right side of Eq.~(\ref{eq:kin2}) behaves as $\partial_t \mathcal{A}/(\beta \omega)$, 
the kinetic equation will be approximated by 
\begin{equation}
\label{eq:wt0}
 \partial_t \mathcal{A}(k,t)= \beta \omega(k) K_l[\mathcal{A}] (k, t)  ,  \quad k \rightarrow 0 . 
 \end{equation}

The integral operator that we have obtained in Eq.~(\ref{eq:asymp0}) is slightly different from
 those that appear in the linearization of  the kinetic equation of wave turbulence. 
 However, by assuming  the validity of integration by parts  
 and noting that  $n_0 \sim (\beta \omega(k))^{-1}$  at low momentum, 
 the kinetic equation~(\ref{eq:kin2})  adopts the form
 \begin{equation}
 \label{eq:wt1}
 \partial_t \mathcal{A}(k,t)= \beta \omega(k) K_l[\mathcal{A}] (k, t) = 
 k^{-h} 
 \int_0^\infty \mathcal{V}_l\left( \frac{k}{q}\right)  \mathcal{A}(q,t) \frac{dq}{q}  ,  \quad k \rightarrow 0 , 
 \end{equation}
 where $h=1$. 
The kernel $\mathcal{V}_l(x)$ may include a term proportional  to $\delta(x-1)$,  
but in our case this does not occur. 
Essentially,  such an equation is the final result of the linearization 
procedure of the nonlinear equation of wave turbulence~\cite{book}, 
which may be obtained from the original~(\ref{eq:kin})
by keeping the leading part of the integrand when $n \gg 1$~\cite{Son}.
Nevertheless, in the absence of the large momentum part of the distribution functions  assuring  
the fall off of the integrand at large momentum, 
the naive justification of this procedure may be inappropriate 
and a more delicate approach is required. 
That is  the reason why we have addressed the linearization of  the full quantum kinetic equation 
preserving suitable integrability properties. 
 
In the remainder of this section we shall focus on the solutions of Eq.~(\ref{eq:wt0}) with $K_l$ approximated by 
Eq.~(\ref{eq:asymp0}). 
 

\subsection{Evolution of perturbations close to $T_\mathrm{c}$}

In order to obtain the characteristic time scale of the kinetic equation, 
it is convenient to introduce the dimensionless momentum  variable $\overline{k} = k \sqrt{\beta/m}$; 
the  scale $(m/\beta)^{1/2}$ corresponds to the average value of  the momentum carried by a quasiparticle in the 
weakly interacting gas.  
It turns out that the right side of Eq.~(\ref{eq:wt0}) reads 
\begin{equation}
\label{eq:ectau}
\partial_t \mathcal{A}(\overline{k}, t)=
\frac{1}{\tau} \int_0^\infty \mathcal{H}_l\left( \frac{k}{q}\right) \partial_{\overline{q}} \mathcal{A}(\overline{q},t) \frac{dq}{q}  - 
\frac{\mathcal{I}_l}{\tau \overline{k}}  \mathcal{A}(\overline{k},t) , 
\end{equation}
where  $\tau^{-1}$ 
sets the inverse time scale corresponding to  the scattering rate 
involving three excitations in the presence of the (small) condensate. 
It is given by 
\begin{equation}
\frac{1}{\tau} =  \frac{32 \pi n_\mathrm{c} a^2}{\sqrt{\beta m}}.  
\end{equation} 
Here $\beta$ may be replaced by the inverse critical temperature of the ideal gas 
$\beta_\mathrm{c} =  m \zeta(3/2)^{2/3}/(2\pi n^{2/3})$ since the above expression is linear in 
the small depletion $n_\mathrm{c}/n$.  
We set $\overline{t}=t/\tau$ for the reduced variable time. 
The explicit form of 
$\mathcal{H}_0(x)$  is derived in the Appendix~\ref{sec:app} 
(the computations for the cases $l=1,2$ proceed in an analogous way).
The results for $l=0,1,2$ are
\begin{eqnarray}
\label{eq:h0}
 \mathcal{H}_0(x) &=& \theta(1-x) \frac{1}{x} \ln \left(\frac{1+x^2}{1-x^2} \right)  + 
            \theta(x-1) \frac{1}{x} \ln \left(1- \frac{1}{x^4} \right) , \\ 
 \mathcal{H}_1(x) &=& \theta(1-x) \frac{1}{x} \ln \left(\frac{1+x}{1-x} \right)  +
  \theta(x-1)\left[  \frac{2}{x^2} -\frac{1}{x} \ln \left(\frac{x+1}{x-1} \right)\right]  ,    \\ 
 \mathcal{H}_2(x) &=&  -\theta(1-x) \frac{1}{2x} 
   \ln \left[(1+x^2)(1-x^2)^2 \right] \nonumber \\ 
   && +
  \theta(x-1)\left[  \frac{3}{2 x^3} -\frac{1}{2 x} \ln \left(\frac{x^4+x^2}{(x^2-1)^2} \right)\right]     , \\
   \mathcal{I}_0 &=& 0, \\ 
   \mathcal{I}_1 &=& 2 - \ln 16, \\ 
   \label{eq:i2}
   \mathcal{I}_2 &=& \frac{3}{2} - \frac{1}{2} \ln 256 , 
\end{eqnarray} 
Hereafter in this section we will drop the lines over the reduced momentum variables.  

The most efficient way to manage the convolution integral  of the kinetic equation is to use the Mellin transform with respect to momentum 
and the Laplace transform with respect to time. 
If $\mathcal{F}(s,\lambda)$ denotes the image of $\mathcal{A}(k, \overline{t})$,  
\begin{equation}
 \mathcal{F}(s,\lambda) = 
  \int_0^\infty d\overline{t}\, e^{-\lambda \overline{t}}  \int_0^\infty \mathcal{A}(k, \overline{t}) k^{s-1} dk ,  
\end{equation}
the evolution equation becomes
\begin{equation}
\lambda \mathcal{F}(s,\lambda) = W_{\mathcal{H}_l}(s) \left[ -(s-1) \mathcal{F}(s-1,\lambda) \right]  
-\mathcal{I}_l \mathcal{F}(s-1,\lambda) + 
\Psi(s), 
\end{equation}
or alternatively
\begin{equation}
\label{eq:mell0}
\lambda \mathcal{F}(s+1,\lambda) = W_{\mathcal{V}_l}(s)  \mathcal{F}(s,\lambda)   +
\Psi(s +1), 
\end{equation}
where
\begin{equation}
W_{\mathcal{V}_l}(s) \equiv -s W_{\mathcal{H}_l}(s+1) - \mathcal{I}_l. 
\end{equation}
Here $W_{\mathcal{H}_l}(s)$  and $\Psi(s)$ are the Mellin transforms of  $\mathcal{H}_l(x)$ and 
the initial condition $\mathcal{A}(k, 0)$, while 
$W_{\mathcal{V}_l}(s)$ denotes the Mellin image of the kernel $\mathcal{V}_l(x)$ in Eq.~(\ref{eq:wt1}). 
We will assume that $\mathcal{A}(k, 0)$ has compact support to assure that $\Psi(s)$ is an entire function. 
Let us first concentrate in the $l=0$ case. 
From  Eq.~(\ref{eq:h0})  we see that the Mellin function is given by 
\begin{equation}
W_{\mathcal{V}_0}(s) = -2 \gamma_\mathrm{E} -2 \psi\left(\frac{s}{2} \right)-\pi \cot \left( \frac{\pi s}{4}\right) , 
\quad \mathrm{Re}\,  s \in (-2,4) ,
\end{equation}
where $\gamma_\mathrm{E}$ is the Euler constant and $\psi(z)= \Gamma'(z)/\Gamma(z)$ is the digamma function. 
Note the reflection property $W_{\mathcal{V}_0}(1+s)=W_{\mathcal{V}_0}(1-s)$, 
in accordance with the symmetry of $\mathcal{U}_0(k, k')$. 
The fundamental strip where $W_{\mathcal{V}_0}(s)$ is  analytic is $\mathrm{Re}\,  s \in (-2, 4)$, and contains two simple zeros at $s=0, 2$.  
It is important for what follows to introduce the winding number of the Mellin function $W_{\mathcal{V}_l}(s)$ as 
 the increment  divided by $2\pi$  of  the argument  of $W_{\mathcal{V}_l}(s)$   as 
$s$ moves from $\sigma - i\infty$ to $\sigma + i\infty$ along the line $\mathrm{Re}\,  s  = \sigma$. 
In the $l=0$ case, 
one can check numerically  that the winding number $\kappa(\sigma) $ of the Mellin function vanishes when
$ \sigma \in (0, 2)$. 

To proceed further one must solve Eq.~(\ref{eq:mell0}) for $ \mathcal{F}(s,\lambda)$.
(In this paragraph we will assume an arbitrary non-negative value for $h$.)
This can be accomplished by following the method of Ref.~\cite{Balk}, 
briefly  reviewed in the sequel.  
These authors have shown that when an interval of zero winding exists 
 within the fundamental strip,  $(\sigma_{-},\sigma_{+}) \subset  (a,b)$, 
then the equation~(\ref{eq:mell0}) has a unique solution. 
It can be constructed using a particular solution $\mathcal{B}(s)$ of the homogeneous equation
\begin{equation}
\mathcal{B}(s + h) = -W(s) \mathcal{B}(s), \quad \mathrm{Re}\,  s \in (a, b) , 
\end{equation}
whose requisite properties have been described in Ref.~\cite{Balk}.
The most important features of $\mathcal{B}(s)$
are the meromorphic property in 
the strip  $\mathrm{Re}\,  s \in (a,b + h)$, 
and its analyticity and absence of zeros 
in the strip $\mathrm{Re}\,  s \in (\sigma_{-},\sigma_{+} + h)$. 
The location of the poles and zeros  of $\mathcal{B}(s)$ outside 
that interval is dictated by 
the set of zeros  of the Mellin function within the fundamental strip.  
They determine the asymptotic properties of the solution. 
In particular,  it turns out  that the pole of $\mathcal{B}(s)$ 
with maximal real part determines the asymptotics of   $\mathcal{A}(k, \overline{t})$
as $k \rightarrow 0$.  
Now, by expressing the Mellin image of the perturbation as
\begin{equation}
\label{eq:conv}
\mathcal{F}(s,\lambda) =  \mathcal{B}(s) f(s, \lambda), 
\end{equation} 
the function $f$ satisfies the inhomogeneous difference equation with constant coefficients 
\begin{equation}
\lambda f(s,\lambda) + f(s-h,\lambda)  = \frac{\Psi(s)}{ \mathcal{B}(s)},  
\quad \mathrm{Re}\,  s \in (\sigma_-, \sigma_+ +h)
\end{equation} 
which has an  inverse Mellin image given by 
\begin{eqnarray}
\label{eq:ak}
(\lambda  + k^{-h} ) a(k,\lambda)  &=& Q(k)  \nonumber \\ 
&=&
\frac{1}{2\pi i} \int_{\sigma - i \infty}^{\sigma + i \infty} \frac{\Psi(s)}{ \mathcal{B}(s)} k^{-s} ds,  
 \quad \sigma  \in (\sigma_{-}, \sigma_{+}+h) . 
\end{eqnarray} 
The end step is to write the convolution corresponding to Eq.~(\ref{eq:conv}) and to invert
 the Laplace transform. This yields the solution
\begin{equation}
\label{eq:sol}
\mathcal{A}(k, \overline{t}) =  \int_0^\infty \mathcal{Z} \left( \frac{k}{q}\right)  
 \exp\left(-q^{-h}\,  \overline{t} \right)Q(q) \frac{dq}{q} , 
\end{equation} 
in terms of  
the inverse Mellin transform of the base function 
\begin{equation}
\label{eq:ZZ}
\mathcal{Z}(k) = \frac{1}{2\pi i} \int_{\sigma - i \infty}^{\sigma + i \infty} \mathcal{B}(s) k^{-s} ds, 
\quad \mathrm{Re}\,  s \in (\sigma_{-}, \sigma_{+}+h) . 
\end{equation}
Notice that  when $h>0$ the asymptotic expansion  of the solution~(\ref{eq:sol}) 
as $\overline{t}  \rightarrow \infty$ requires
the determination of the leading behavior of $Q(k)$  as $k \rightarrow \infty$.   
According to Eq.~(\ref{eq:ak}), this comes from the zero of 
$\mathcal{B}(s)$ with minimal real part. 

Next we turn to the $l=0$ case.
Due to the presence of the zero  of $W_{\mathcal{V}_0}(s)$ at $s=0$, 
the function $\mathcal{B}_0(s)$ has  a sequence of simple poles  at $s=0,-1,-2,\ldots$, 
with the behavior  
\begin{equation}
\mathcal{B}(s) \sim -\frac{\mathcal{B}(1)}{W_{\mathcal{V}_0}'(0)} \frac{1}{s}, \quad    s \rightarrow 0 , 
\end{equation}
then 
\begin{equation}
\label{eq:Zk}
\mathcal{Z}(k) \sim -\frac{\mathcal{B}(1)}{W_{\mathcal{V}_0}'(0)}  , \quad    k \rightarrow 0 , 
\end{equation}
where $\mathcal{B}(1)$ is an arbitrary constant. 
This leads to a behavior independent of  $k$ at low momentum 
\begin{equation}
\label{eq:ask0}
\mathcal{A}(k, \overline{t})  \sim -\frac{\mathcal{B}(1)}{W_{\mathcal{V}_0}'(0)} 
  \int_0^\infty   
 \exp\left(-\overline{t}/q \right) Q(q) \frac{dq}{q} ,  \quad k \rightarrow 0 .   
\end{equation} 
Note that $\mathcal{B}(1)$  must cancel  another similar factor arising from $Q(q)$. 
On the other hand, the zero of $\mathcal{B}(s)$ with minimal real part 
is at $s=3$, and it comes from the zero of $W_{\mathcal{V}_0}(s)$ at $s=2$.  
Using $\mathcal{B}(s+2) = W_{\mathcal{V}_0}(s+1) W_{\mathcal{V}_0}(s) \mathcal{B}(s)$ 
when $s \rightarrow 1$ we obtain 
\begin{equation}
\mathcal{B}(s) \sim  \mathcal{B}(1) W_{\mathcal{V}_0} (1) W_{\mathcal{V}_0}'(2) (s-3) ,  
\quad  s \rightarrow 3 . 
\end{equation}
This zero determines the high momentum limit of $Q(k)$ in the form 
\begin{equation}
\label{eq:Qk}
Q(k) \sim -\frac{\Psi(s=3)}{\mathcal{B}(1) W_{\mathcal{V}_0} (1) W_{\mathcal{V}_0}'(2)} \frac{1}{k^3}, 
\quad k \rightarrow \infty , 
\end{equation} 
so,  the long time behavior  of the solution is 
given by  
 \begin{equation}
 \label{eq:ast}
\mathcal{A}(k, \overline{t}) \sim -\frac{\Psi(s=3)}{\mathcal{B}(1) W_{\mathcal{V}_0} (1) W_{\mathcal{V}_0}'(2)} 
 \int_0^\infty \mathcal{Z} \left( \frac{k}{q}\right)  
 \exp\left(-\overline{t}/q \right) \frac{dq}{q^4} ,  \quad \overline{t} \rightarrow \infty , 
\end{equation} 
which possess the self-similarity property
\begin{equation}
\mathcal{A}(k, \overline{t} v) = \frac{1} {\overline{t}^{3}} \mathcal{A}\left(\frac{k}{\overline{t}},  v\right) . 
\end{equation}
Finally, it is easy to derive in closed form the asymptotics in the 
more restricted regime $k \rightarrow 0, \overline{t} \rightarrow \infty$ either from 
Eq.~(\ref{eq:ask0}) combined with~(\ref{eq:Qk}),  
or from Eq.~(\ref{eq:ast}) combined with~(\ref{eq:Zk}). The result is 
 \begin{equation}
\mathcal{A}(k, \overline{t}) \sim 
\frac{2 \Psi(s=3)}{W_{\mathcal{V}_0}'(0) W_{\mathcal{V}_0} (1) W_{\mathcal{V}_0}'(2)}
 \frac{1}{ \overline{t}^3},  \quad k \rightarrow 0,\, \overline{t} \rightarrow \infty , 
\end{equation} 
where 
\begin{equation}
W_{\mathcal{V}_0}'(0) W_{\mathcal{V}_0} (1) W_{\mathcal{V}_0}'(2) = 
\frac{\pi^4}{144} \left( \pi - \ln 16 \right) . 
\end{equation}

The source term $\delta \Gamma_{12}$  only receives contribution 
from the $l=0$ component, 
\begin{eqnarray}
\delta \Gamma_{12}  &=& \frac{d}{d t} \int n_0 (1+n_0) \chi(\bm{k}, t) \frac{d^3 \bm{k}}{(2\pi)^3} \nonumber  \\
   &=&  \frac{1}{4 \pi^{5/2}} \frac{d}{d t} \int_0^\infty n_0 (1+n_0) \beta \omega(k) \mathcal{A}_{00}(k, t)k^2 dk, 
\end{eqnarray} 
or using  the above low  momentum approximation 
\begin{eqnarray}
\label{eq:source}
\delta \Gamma_{12}  &\sim & \frac{1}{4 \pi^{5/2}} \frac{d}{d t} \int_0^\infty  
\frac{\mathcal{A}_{00}(k, t)}{\beta \omega(k)} k^2 dk \nonumber \\
      &=& \frac{m^{3/2}}{2 \pi^{5/2} \beta^{3/2} \tau}   \frac{d}{d \overline{t}}  \mathcal{F}(s=1,\overline{t}) , 
\end{eqnarray}
which contains the partial Mellin transform of $\mathcal{A}_{00}(k, \overline{t})$ for $s=1$.  
From the convolution in Eq.~(\ref{eq:sol}) one finds
\begin{equation}
\mathcal{F}(s,\overline{t}) = \mathcal{B}(s) \int_0^\infty   
 \exp\left(-\overline{t}/q \right) Q(q) q^{s-1} dq .  
\end{equation} 
The substitution of the large momentum behavior of $Q(k)$ given in Eq.~(\ref{eq:Qk})
yields the asymptotics
\begin{equation}
\mathcal{F}(s,\overline{t}) \sim -\mathcal{B}(s) \frac{\Psi(3) \Gamma(3-s) \overline{t}^{s-3} }
{\mathcal{B}(1) W_{\mathcal{V}_0} (1) W_{\mathcal{V}_0}'(2)}  , 
\quad \overline{t} \to \infty, \quad \mathrm{Re}\, s < 3.
\end{equation} 
Combining this with Eq.~(\ref{eq:source}) then gives 
the long-time behavior  of the source term 
\begin{equation}
\delta \Gamma_{12}  \sim \frac{m^{3/2} \Psi(3)}{ \pi^{5/2} \beta^{3/2} \tau   W_{\mathcal{V}_0} (1) W_{\mathcal{V}_0}'(2)}
\frac{1}{ \overline{t}^3}, 
\quad  \overline{t} \rightarrow \infty . 
\end{equation} 
Since the energy of the perturbation is proportional to $\mathcal{F}(s=3,\overline{t})$, we see that 
\begin{equation}
\mathcal{F}(s=3,\overline{t} ) = \Psi(3), \quad  \overline{t} \rightarrow \infty, 
\end{equation}
in accordance with energy conservation. 

What about the non-isotropic perturbations? The Mellin functions  corresponding to  the $l=1, 2$ harmonics are
\begin{eqnarray}
W_{\mathcal{V}_1}(s) &=& -2 \gamma_\mathrm{E} + \frac{2}{s-1} - 
  \psi\left(\frac{1+s}{2} \right)-\psi\left(\frac{1-s}{2} \right) , 
  \;  \mathrm{Re}\,  s \in (-1,3) , \\ 
W_{\mathcal{V}_2}(s) &=& -2 \gamma_\mathrm{E} + \ln 16 + \frac{6}{s(s-2)} \nonumber \\  
 &&-\pi \cot\left(\frac{\pi s}{2}  \right) + 
 \frac{\pi}{2 \sin\left(\frac{\pi s}{2}  \right)}  - 2 \psi\left(\frac{s}{2} \right) , 
  \quad  \mathrm{Re}\,  s \in (-2, 4) . 
\end{eqnarray}
Now $W_{\mathcal{V}_1}(s)$ has a double zero at $s=1$,  and 
the winding number $\kappa$ does not vanish on the entire strip of analyticity; 
the corresponding values are $1, (-1)$ if  $\mathrm{Re}\,  s \in (1, 3)$,  ($\mathrm{Re}\,  s \in (-1, 1)$).  
For $l=2$, the winding number  does not assume the value zero either, but now  
$\kappa=1$ on the entire interval (-2,4), and  the Mellin function does not vanish in 
the strip of analyticity. 
In these situations of absence of an interval of zero winding, Balk and Zakharov~\cite{Balk} 
have shown that the initial value problem for $\mathcal{A}(k, 0)$ is not well posed, 
signaling the non-uniqueness of the solution when  $\kappa >0$ on the entire interval (as in  the $l=2$ case), 
or a strong instability  of exponential growth  if $\kappa\gtrless  0$ (as  in the $l=1$ case).
In terms of the Mellin function, we have $W_{\mathcal{V}_{1,2}}(0) > 0$,  which at least  is 
a sufficient condition~\cite{Balk} for the instability of the Rayleigh-Jeans spectrum $n\propto k^{-2}$  under 
anisotropic perturbations.  
Thus, it seems praiseworthy to assign this property to the Bose-Einstein equilibrium solution  of the 
kinetic equation with a collision term given by Eqs.~(\ref{eq:crit1}) and~(\ref{eq:crit2}). 

Finally, it may be instructive to compare these findings with those of Ref.~\cite{Valle}. 
The application of the methods of wave turbulence theory to the study of  the
binary collision term  
$C_{22}$,  approximated when $n_0 \gg 1$,  
 showed the Mellin function $W_{22}(s)$ to be symmetric with respect to $s=3/2$, 
 reflecting the symmetry of the corresponding kernel $\mathcal{U}_0$.  
 At the same time, it was shown that 
 $W_{22}(s = 0) > 0$ while  $W_{22}(s = 1)=0$.
As a consequence,  such approximated form of $C_{22}$ destroyed energy conservation 
but was consistent with particle number conservation. 
By simply shifting the minimum of $W_{22}(s)$ one can get $W_{22}(s = 0) = 0$ and   
$W_{22}(s = 1)\neq 0$. 
So that 
another approximated form of  $C_{22}$, 
which conserves the energy but destroys the conservation of the particle number, 
appears as possible. 
However, 
given the reflection property of $W_{22}(s)$ and the location of its zeros, 
it does not seem possible to get a low momentum approximation 
maintaining both conservation laws. 


\section{\label{sec:class}
 Evolution of perturbations in the thermal regime $\beta c k \ll 1$}

\begin{figure}
\centering
    \includegraphics{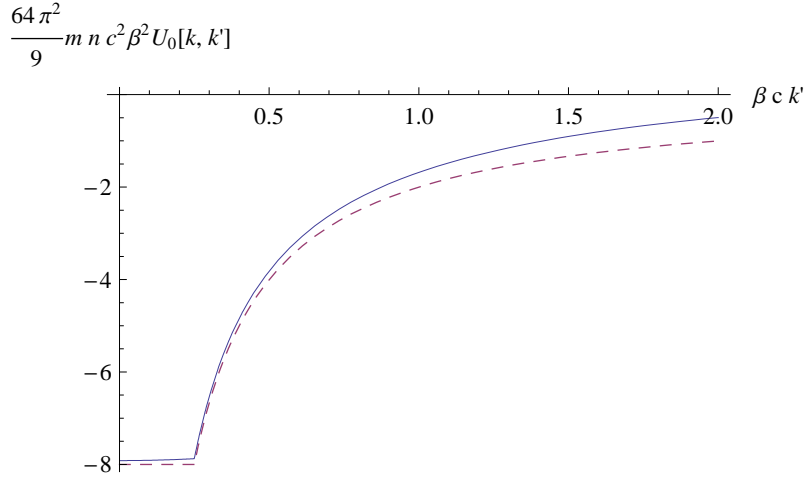}
    \caption{\label{fig:U0class} 
    The kernel $U_0(k, k') $ 
    in the low temperature regime when $\beta c k \ll  1$. 
    The continuous  line corresponds to the exact kernel for $\beta c k=1/4$. 
    The dashed line is the kernel obtained with $\mathcal{U}(k,k')$ 
    approximated by Eq.~(\ref{eq:class}) for  $\beta c k=1/4$.} 
\end{figure}
 
In this section  we consider the  regime $\omega(k) \ll k_\mathrm{B} T \ll g n$, 
where the energy of quasiparticles is $\omega = c k$, with $c = \sqrt{g n m^{-1}}$.  
Their low-energy interactions are well described by the squared amplitude 
$|\mathcal{M}|^2$ given by Eq.~(\ref{eq:low2}). 
By ignoring the next  positive contribution $(8 m^2 c)^{-1} k^3$ in the dispersion law 
all  $\mathcal{U}_l$ become equal to  $(2l+1)\mathcal{U}_0$,  due to the 
collinearity of wave vectors in the collision integral. 
Therefore in this approximation, the kinetic equation is the same for all values of $l$. 
The low momentum limit of this kernel  is  
\begin{equation}
\label{eq:class}
\mathcal{U}_0(k, k') \sim -\frac{9}{32\pi^2 m n \beta^3 c^3} \left(\frac{\theta(k-k')}{k} + \frac{\theta(k'-k)}{k'}  \right), 
\quad \beta c k \to 0, \; \beta c k' \to 0 .
\end{equation} 
A comparison between the exact and the approximated kernels is shown in Fig.~\ref{fig:U0class}.
Our first task is to find an evolution equation consistent with energy conservation. 
When Eq.~(\ref{eq:class}) is substituted into Eq.~(\ref{eq:fred}), 
the latter turns into 
\begin{eqnarray}
K[\mathcal{A}](k,t) &=& -\frac{9 k^3 }{8\pi m n \beta^3 c^3}\int_0^\infty 
\left( \frac{k'^4}{k^4} \theta(k-k') + \frac{k'^3}{k^3} \theta(k'-k) \right)  \mathcal{A}(k',t) \frac{dk'}{k'} \nonumber  \\ 
 &&+\frac{9 k^3 }{8\pi m n \beta^3 c^3} \left(\frac{\Lambda^3}{3 k^3}-\frac{1}{12} \right) \mathcal{A}(k,t) , 
\end{eqnarray}
where we have introduced an ultraviolet cutoff $\Lambda \gg k$ in the integral which gives
the coefficient of $\mathcal{A}(k,t)$.
We left for the moment this  term 
undetermined. 
The well-defined convolution term in the above equation 
determines the (provisional) class of admissible solutions 
of the approximated evolution equation.  
Such a class would be  
formed by functions $\mathcal{A}(k,t)$ which grow as $k \to 0$ more slowly than $k^{-4}$, 
and  fall off 
faster than  $k^{-3}$ as $k \to \infty$. 
In accordance with that, the integral for the Mellin transform of the gain term 
\begin{equation}
-\int_0^\infty \left[ x^{-4} \theta(x-1) +  x^{-3} \theta(1-x)\right]  x^{s-1} dx = \frac{1}{(s-3)(s-4)}, 
\end{equation}
is convergent for $3 < \mathrm{Re} s  < 4$. 
Note however that the analytic continuation of this integral  to $s=0$ 
is  the value of the gain term formally evaluated for 
$\mathcal{A}(k',t)=1$. 
Based on energy conservation, we may insist on considering  
the functions $\mathcal{A}= \mathrm{constant}$
within the admissible  class of solutions, so the loss term must 
cancel out this value $1/12$. 
Therefore, the replacement $\Lambda^3/(3 k^3) - 1/12 \rightarrow -1/12$ 
produces such a cancellation. 

The approximated evolution equation obtained in this way reads
\begin{eqnarray}
\label{eq:bad}
\partial_t \mathcal{A}(k, t) &=& -\frac{9 k^4 }{8\pi m n \beta^2 c^2}\int_0^\infty 
\left( \frac{k'^4}{k^4} \theta(k-k') + \frac{k'^3}{k^3} \theta(k'-k) \right)  \mathcal{A}(k',t) \frac{dk'}{k'}  \nonumber  \\ 
 &&  -\frac{3 k^4}{32\pi m n \beta^2 c^2} \mathcal{A}(k, t) , 
\end{eqnarray}
which again has the form of a linearized kinetic equation of wave turbulence, 
where the degree of homogeneity is now $h=-4$. 
In fact, we will find that the above equation still 
introduces a conflict with energy conservation. 
This defect is repaired by adding a source term independent of $k$ 
to the right-hand side of Eq.~(\ref{eq:bad}). It has the form 
\begin{equation}
\Phi(t) = \frac{9}{8\pi m n \beta^2 c^2}\int_0^\infty  \mathcal{A}(k',t) k'^3  dk' , 
\end{equation}
and  produces a partial cancellation of the integral term of Eq.~(\ref{eq:bad}) 
with the result
\begin{eqnarray}
\label{eq:good}
\partial_t \mathcal{A}(k, t) &=& -\frac{3 k^4}{32\pi m n \beta^2 c^2} \mathcal{A}(k, t)  \nonumber  \\ 
 && +\frac{9 k^4 }{8\pi m n \beta^2 c^2}\int_k^\infty 
\left( \frac{k'^4}{k^4}- \frac{k'^3}{k^3}  \right)  \mathcal{A}(k',t) \frac{dk'}{k'} .
\end{eqnarray}
This implies that the class of solutions of the improved equation must be  changed. 
Such a class is formed indeed by functions which fall off faster than $k^{-4}$  as $k \to \infty$. 
In order to understand the role of the source term and how it arises, 
it is convenient first to discuss the solution of 
the wrong evolution based on Eq.~(\ref{eq:bad}). 

By introducing the function 
\begin{equation}
\mathcal{V}(x) = -\delta(x-1) - 12 x^{-4} \theta(x-1) - 12 x^{-3} \theta(1-x), 
\end{equation}
we define the Mellin function corresponding to  Eq.~(\ref{eq:bad}) as 
\begin{equation}
W_{\mathcal{V}}(s) =\int_0^\infty  \mathcal{V}(x) x^{s-1} dx = -\frac{s(s-7)}{(s-3)(s-4)} , 
 \quad \mathrm{Re}\,  s \in (3, 4) ,
 \end{equation} 
which has  zero winding  in the entire  strip of analyticity, and 
$W_{\mathcal{V}}(s+7/2) = W_{\mathcal{V}}(7/2-s)$. 
The corresponding difference equation for the Laplace-Mellin image of $\mathcal{A}$  becomes 
\begin{equation}
\label{eq:mell1}
\lambda \mathcal{F}(s-4,\lambda) = v W_{\mathcal{V}}(s)  \mathcal{F}(s,\lambda)   +
\Psi(s - 4),  \quad \mathrm{Re}\,  s \in (3, 4), 
\end{equation}
where $v = 3/(32 \pi m n \beta^2 c^2)$. 
The method used to solve  Eq.~(\ref{eq:mell1}) is the same as before. 
We seek an appropriate solution  of $\mathcal{B}(s-4) = -W_{\mathcal{V}}(s) \mathcal{B}(s)$
for  $\mathrm{Re}\,  s \in (3, 4)$,
and we write the solution in the factorized form   
\begin{eqnarray}
\mathcal{F}(s, \lambda)&=& \mathcal{B}(s)\int_0^\infty \frac{Q(k)}{v k^4+ \lambda}  k^{s-1} dk  ,  \\ 
 Q(k) &=& 
\frac{1}{2\pi i} \int_{\mathrm{Re}\,  s- i \infty}^{\mathrm{Re}\,  s + i \infty} \frac{\Psi(s)}{ \mathcal{B}(s)} k^{-s} ds .
\end{eqnarray}
In view of Eq.~(\ref{eq:mell1}) and such a factorization, 
we need a solution $\mathcal{B}(s)$ which has neither zeros nor poles 
when $\mathrm{Re}\,  s \in (-1, 4)$. 
We can write the required function as 
\begin{equation}
\mathcal{B}(s) = \frac{\Gamma\left(\frac{7-s}{4} \right) \Gamma\left(1+\frac{s-3}{4} \right)}
 {\Gamma\left(\frac{4-s}{4} \right) \Gamma\left(1+\frac{s}{4} \right)} 
 =\frac{3-s}{s}\frac{\sin \left(\frac{\pi s}{4} \right)}{\sin \left(\frac{\pi}{4} ( s+1)\right)} . 
\end{equation}
The other solutions of the homogeneous difference equation 
in terms of products and fractions of $\Gamma$-functions,  such as 
\begin{eqnarray}
 \frac{\Gamma\left(-\frac{s}{4} \right) \Gamma\left(1+\frac{s-4}{4} \right)}
 {\Gamma\left(\frac{3-s}{4} \right) \Gamma\left(1+\frac{s-7}{4} \right)}
 &=& \frac{3-s}{s}\frac{\sin \left(\frac{\pi}{4} ( 1+s)\right)}{\sin \left(\frac{\pi}{4} s \right)} ,  \\
 \label{eq:btrue}
 \frac{\Gamma\left(\frac{7-s}{4} \right) \Gamma\left(-\frac{s}{4} \right)}
 {\Gamma\left(\frac{3-s}{4} \right) \Gamma\left(\frac{4-s}{4} \right)}  
 &=& \frac{\Gamma\left(1+\frac{s-3}{4}\right) \Gamma\left(1+\frac{s-4}{4} \right)}
 {\Gamma\left(1+\frac{s-7}{4} \right) \Gamma\left(1+\frac{s}{4} \right)}
 = \frac{s-3}{s} ,  
 \end{eqnarray} 
are not appropriate for the present problem since
 they have a pole at $s=0$ within the strip $\mathrm{Re}\,  s \in (-1, 4)$.
In accordance with the factorization of $\mathcal{F}(s, \lambda)$, 
one finds the solution as the convolution 
\begin{equation}
\mathcal{A}(k,t) =  \int_0^\infty \mathcal{Z} \left( \frac{k}{q}\right)  
 \exp\left(-v q^{4} t\right)Q(q) \frac{dq}{q} , 
\end{equation} 
where $\mathcal{Z}(k)$ and $Q(k)$ are given by the integrations in 
Eqs.~(\ref{eq:ZZ}) and~(\ref{eq:ak})  with $\sigma \in (-1, 4)$.  
Thus, to derive the asymptotics of the perturbation as $t \rightarrow \infty$, 
we must evaluate the low momentum behavior of $Q(k)$. 
This is dictated by the pole of $\mathcal{B}(s)^{-1}$ at $s=-4$ which leads to 
\begin{equation}
\label{eq:fin2}
Q(k) \sim- \frac{2^{7/2} \Psi(-4)}{7 \pi} k^4 , \quad k \rightarrow 0 . 
\end{equation}
If $\mathcal{F}(s, t)$ denotes the Mellin image  of the solution $\mathcal{A}(k, t)$, 
the above  result permits the calculation of the leading behavior  of the quotient  
$\mathcal{F}(s, t)/\mathcal{B}(s)$ as $t \rightarrow \infty$. 
It reads 
\begin{eqnarray}
\label{eq:partial} 
\frac{\mathcal{F}(s, t)}{\mathcal{B}(s)}&=& \int_0^\infty \exp(-v k^4 t) Q(k)  k^{s-1} dk \nonumber \\ 
 &\sim& -\frac{2^{3/2} \Psi(-4)}{7\pi} \Gamma\left(1+\frac{s}{4}\right) (v t)^{-1-s/4}, 
\quad t  \rightarrow \infty, \;  \mathrm{Re}\,  s > -4 . 
\end{eqnarray}
We can see now the violation of energy conservation. 
In the approximation where $n_0 \gg 1$ this requires for any time that
\begin{eqnarray}
\frac{dE}{dt}  &=& \frac{1}{4 \pi^{5/2} \beta} \frac{d}{d t} \int_0^\infty  
\mathcal{A}_{00}(k, t) k^2 dk \nonumber \\
      &=& \frac{1}{4 \pi^{5/2} \beta}   \frac{d}{dt}  \mathcal{F}(s=3,t) = 0 , 
\end{eqnarray}
or $\mathcal{F}(s=3,t)= \mathcal{F}(s=3,t=0)=\Psi(3)$, 
since $\Psi(3)$ corresponds to the energy of the initial perturbation. 
 But, in accordance with the above asymptotics, 
 the quantity $\mathcal{F}(s=3,t)$ assumes a value proportional to 
 $\Psi(-4) \mathcal{B}(3) t^{-7/4}$ for large time, so $dE/dt  \neq 0$. 
 Another indication that Eq.~(\ref{eq:bad})  is  conflicting  
 comes from 
the fact that, in order to check energy conservation,
the integration of  both sides after multiplication by 
 $k^2$ leads to 
\begin{equation}
\frac{d}{dt} \int_0^\infty \mathcal{A}(k,t) k^2 dk =  
v  \int_0^\infty  \mathcal{A}(q,t) q^6 dq \int_0^\infty \mathcal{V}(x) x^6 dx,  
\end{equation} 
which does not converge for the class of functions considered up till now. 

To  obtain an improved evolution equation, 
let us consider a factorized solution $\mathcal{F}(s, t)$ 
for a different base function respecting energy conservation. 
If we assume that $Q(k) \sim D k^\alpha$  as $k \to 0$, 
then  the limit  behavior of  $\mathcal{F}(s,t)$ as $t \to \infty$ is 
\begin{eqnarray}
\label{eq:partial0} 
\mathcal{F}(s, t)&=& \mathcal{B}(s)\int_0^\infty \exp(-v k^4t) Q(k)  k^{s-1} dk  \nonumber  \\ 
 &\sim&  \frac{D}{4} \mathcal{B}(s)\Gamma\left(\frac{s+\alpha}{4}\right) (v t)^{-(s+\alpha)/4} , 
 \quad \mathrm{Re} (s +\alpha) > 0. 
\end{eqnarray}
Since  energy conservation requires that $\mathcal{F}(s = 3, t=\infty) = \Psi(3)$, 
we see that $\alpha=-3$, and $\mathcal{B}(s)$ must  possess a simple zero at $s=3$ in 
order to cancel the pole of the gamma function. 
Furthermore,  to accomplish the low momentum behavior of $Q(k)$, the prescription for 
$\mathrm{Re}\, s$ in 
\begin{equation}
\label{eq:q1}
Q(k) = 
\frac{1}{2\pi i} \int_{\mathrm{Re}\,  s- i \infty}^{\mathrm{Re}\,  s + i \infty} \frac{\Psi(s)}{ \mathcal{B}(s)} k^{-s} ds,
\end{equation} 
and the analog of~(\ref{eq:ZZ}) must be $\mathrm{Re}\, s >3$. 
Clearly,  if the base function assumes the value 
\begin{equation}
\mathcal{B}(s)=\frac{s-3}{s}, \quad \mathrm{Re}\, s > 3, 
\end{equation}
which was discarded before, 
the requirement  $\mathcal{F}(s=3,t)=\Psi(3)$
is accomplished for large time. 

Some features of the solution generated by this base function  are the following. 
The inverse images  of $\Psi(s)/\mathcal{B}(s)$  and $\mathcal{B}(s)$ are simply 
\begin{eqnarray}
Q(k) &=& \mathcal{A}(k,0) +  \frac{3}{k^3} \int_k^\infty \mathcal{A}(q,0) q^2 dq, \\ 
\mathcal{Z}(k) &=& \delta(k-1) -3 \theta(1-k) , 
\end{eqnarray}
and the  behavior of $Q(k)$ as $k \to 0$ reads 
\begin{equation}
\label{eq:lowk}
Q(k) \sim 3 \Psi(3) k^{-3}, \quad k \to 0 .
\end{equation} 
Hence the corresponding solution is written as 
 \begin{eqnarray}
\label{eq:sol2}
\mathcal{A}(k, t) &=&  \int_0^\infty \mathcal{Z} \left( \frac{k}{q}\right)  
 \exp\left(-v q^{4}\,  t \right)Q(q) \frac{dq}{q} , \nonumber \\  
 &=& \exp(-v k^4 t) Q(k)
  - 3 \int_k^\infty \exp(-v q^4 t) Q(q) \frac{dq}{q}, 
\end{eqnarray}
which for large $t$ assumes the value
\begin{equation}
\mathcal{A}(k, t) \sim \exp\left(-v k^{4}\,  t \right)Q(k) , \quad t \to \infty. 
\end{equation}
Similarly, the behavior or the source term as $t \rightarrow \infty$ 
becomes
\begin{eqnarray}
\delta \Gamma_{12}  &= & \frac{1}{4\pi^{5/2} \beta c}  \int_0^\infty  
\partial_{t} \mathcal{A}_{00}(k, t) k\, dk \nonumber \\ 
 &=&  \frac{v}{4\pi^{5/2} \beta c} \left[ -\int_0^\infty   \exp(-v k^4 t) k^5 Q(k) dk  
 +3 \int_0^\infty \frac{dq}{q} \exp(-v q^4 t) q^4 Q(q) \int_0^q k\, dk \right]  \nonumber \\   
&=& \frac{v}{8\pi^{5/2} \beta c} \int_0^\infty   \exp(-v k^4 t) k^5 Q(k) dk ,  
\end{eqnarray}
so the insertion of Eq.~(\ref{eq:lowk})
produces  
\begin{equation} 
\delta \Gamma_{12} \sim \frac{v^{1/4} \Gamma(7/4) \Psi(3)}{8 \pi^{5/2}  \beta c \, t^{3/4}} , 
\quad t \rightarrow \infty . 
\end{equation}
We can check directly energy conservation  for any time, 
\begin{eqnarray}
 \int_0^\infty  
\partial_{t} \mathcal{A}(k, t) k^2 dk  &=& 
  -v \int_0^\infty   \exp(-v k^4 t) k^6 Q(k) dk  \nonumber \\
 &&+3 v \int_0^\infty \frac{dq}{q} \exp(-v q^4 t) q^4 Q(q) \int_0^q k^2 dk = 0. 
\end{eqnarray}

It remains to derive the resulting evolution equation satisfied by $\mathcal{A}(k,t)$  
given in Eq.~(\ref{eq:sol2}). 
 From the homogeneous difference equation evaluated for  
 $\mathcal{B}(s)$ in Eq.~(\ref{eq:btrue})
it follows that 
 \begin{equation}
 W_\mathcal{V}(s) \mathcal{B}(s) = -\mathcal{B}(s-4)= -1 + \frac{3}{s-4}, \quad \mathrm{Re}\,  s \in (3, 4) . 
 \end{equation} 
 Applying the inverse Mellin transform we obtain
 \begin{eqnarray}
 \int_0^\infty \mathcal{V}\left(\frac{k}{q}\right) \mathcal{Z}(q) \frac{dq}{q } &=& 
 -\delta(k-1) -\frac{3}{k^4} \theta(k-1) \nonumber \\
 & =&  -k^{-4}\mathcal{Z}(k) - \frac{3}{k^4}. 
 \end{eqnarray}
 The origin of  the extra term proportional to $k^{-4}$ is the pole of $\mathcal{B}(s)$ at $s=0$. 
 With this result at hand, the integral term of Eq.~(\ref{eq:bad}) 
 evaluated for $\mathcal{A}(k,t)$ in Eq.~(\ref{eq:sol2})  becomes
 \begin{eqnarray}
 \int_0^\infty \mathcal{V}\left(\frac{k}{q}\right) \mathcal{A}(q,t) \frac{dq}{q } &=& \frac{1}{v k^4} \partial_t \mathcal{A}(k,t)  
 -\frac{3}{k^4} \int_0^\infty \exp\left(-v q^4 t \right) Q(q) q^3 dq \nonumber \\ 
 &=& \frac{1}{v k^4} \partial_t \mathcal{A}(k,t)  - \frac{3}{k^4}  \frac{\mathcal{F}(s=4,t)}{\mathcal{B}(s=4)} \nonumber \\
 &=& \frac{1}{v k^4} \partial_t \mathcal{A}(k,t) 
 -\frac{12}{k^4} \int_0^\infty \mathcal{A}(q,t) q^3 dq .
 \end{eqnarray}
The last term can be understood as a self-consistent source which restores energy conservation. 
In consequence, if one uses the partial cancellation produced by this term, 
the corrected evolution equation adopts the form 
\begin{equation}
\partial_t \mathcal{A}(k,t) = v k^4  \int_0^\infty \overline{\mathcal{V}}\left(\frac{k}{q}\right) \mathcal{A}(q,t) \frac{dq}{q}, 
\end{equation}
where 
\begin{eqnarray}
\overline{\mathcal{V}}(x) = -\delta(x-1) -12 x^{-3} \theta(1-x) + 12 x^{-4} \theta(1-x), 
\end{eqnarray}
which gives rise to the same Mellin function as before 
\begin{equation}
W_{\overline{\mathcal{V}}}(s) =  -\frac{s(s-7)}{(s-3)(s-4)}, \quad \mathrm{Re}\, s > 4,  
\end{equation}
but with a different strip of analyticity. 
The relevant interval of zero winding  
is now $\mathrm{Re}\, s \in (7,\infty)$ and 
the base function $\mathcal{B}(s)$ has neither zero nor poles in the corresponding strip  
$\mathrm{Re}\, s \in (3,\infty)$.
We expect  that the evolution equation obtained in this way captures the essential features
of the low momentum regime at low temperature.

\paragraph*{Dominance of $C_{12}$ over $C_{22}$.}
To conclude this section, it is interesting to observe that the collision term  that 
 we have considered in  this and   
the previous section is the prevailing summand in the sum  $C_{12}+C_{22}$ 
in the low momentum regime, $n_0(k) \gg 1$.
This is based on the rules of power counting for the degrees of 
homogeneity $h$ of the linearized collision integral which have been given in Ref.~\cite{Balk}.
These authors have shown that when the dispersion law 
has the form $\omega(k) \propto k^\delta$,  and the 
scattering amplitudes of $C_{12}$ and $C_{22}$ scale according to  
\begin{eqnarray}
\mathcal{M}(\varepsilon \bm{k}, \varepsilon\bm{k}_1,\varepsilon\bm{k}_2) &=&  
\varepsilon^\eta \mathcal{M}( \bm{k}, \bm{k}_1,\bm{k}_2) , \\ 
\mathcal{M}(\varepsilon \bm{k}, \varepsilon\bm{k}_1,\varepsilon\bm{k}_2, \varepsilon \bm{k}_3) &=&  
\varepsilon^\kappa \mathcal{M}( \bm{k}, \bm{k}_1,\bm{k}_2, \bm{k}_3) , 
\end{eqnarray}
the index $h$ 
is given by 
\begin{eqnarray}
h[C_{12}]&=& \delta - 2 \eta - d +  \nu, \\ 
\label{eq:h4}
h[C_{22}]&=& \delta - 2 \kappa - 2 d + 2 \nu,
\end{eqnarray}
where $d$ is the spatial dimension,  and $\nu$ is the exponent in 
a solution $n_0(k) \propto k^{-\nu}$ of $C[n_0]=0$. 

Near the critical temperature, the scattering amplitude  involving four excitations  comes 
from the term $g (\psi^\dagger \psi)^2$ in the effective Hamiltonian, 
where $\psi^\dagger (\psi)$ creates (destroys) excitations, 
while  terms as $g \sqrt{n_\mathrm{c}} \psi^\dagger \psi \psi$ give rise to 
the scattering amplitude  involving three excitations. 
In both cases $\kappa=\eta=0$ since 
$\mathcal{M}_{4,3} \propto g, g \sqrt{n_\mathrm{c}}$. 
The insertion into Eq.~(\ref{eq:h4}) of $\delta=2, \kappa=0, d=3, \nu=2$ produces 
$h[C_{22}]=0$, while $h[C_{12}]=1$. 
Thus, we can neglect the contribution of $C_{22}$ in the linearized description at low momentum. 

At very low temperature, when the excitations are phonons,  
the effective description is made in terms of a Hamiltonian $ H[n(\bm{x}), \theta(\bm{x})]$  depending 
on the particle density $n$  and the Goldstone mode or phonon field $\theta$  conjugate to $n$. 
Now, since each $\theta$-field carries the oscillator  factor $1/\sqrt{\omega(k)}$, the 
conjugate field $n$ goes with $\sqrt{\omega(k)} $. 
Thus,   
a term in the Hamiltonian of order $n^N$ produces a scattering amplitude $\mathcal{M}_N$ 
proportional to  $\prod_i^N \sqrt{k}_i$, where  $k_i$ denotes the momentum 
of the quasiparticle $i$ in the process. This means that $\eta = 3/2$, according to Eq.~(\ref{eq:low2}), 
and $\kappa =  4/2=2$. 
Therefore, with $\delta=1, \nu=1$, we find $h[C_{12}]=-4$ and $h[C_{22}]=-7$, 
which shows that  $C_{12}$ prevails  over $C_{22}$.


\section{\label{sec:qua} Evolution of perturbations in the quantum  regime $\beta c k \gg 1$ }

\begin{figure}
\centering
    \includegraphics{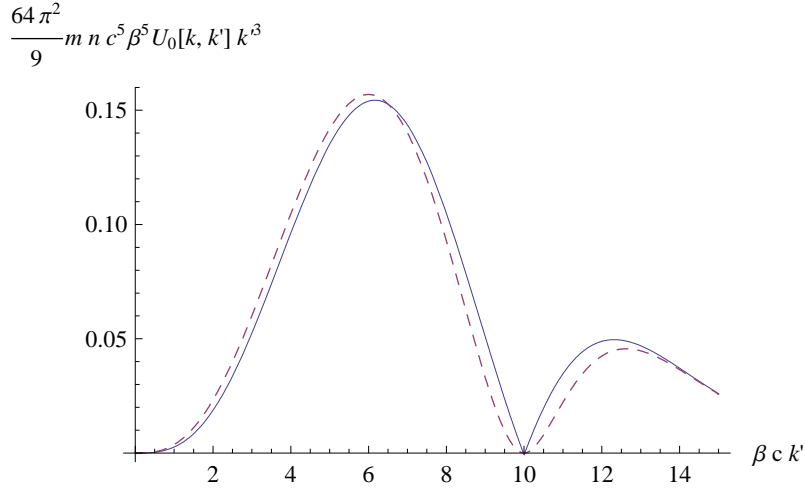}
    \caption{\label{fig:U0quantum}
    $\mathcal{U}_0(k,k') k'^3$ in the quantum regime, $\beta c k \gg  1$. 
    The continuous line corresponds to the exact kernel for $\beta c k=  10$. 
    The dashed line is the kernel obtained with $\mathcal{U}_0(k,k')$ 
    approximated by Eq.~(\ref{eq:quap}) for $\beta c k= 10$.} 
\end{figure}

In this section we consider the kinetic equation and its solutions when the term $\mathcal{U}_0(k,k')$  of  
the collision integral~(\ref{eq:fred}) is approximated in the quantum regime 
$k_\mathrm{B} T \ll c k \ll g n$.
This proceeds by replacing $n_0(1+n_0) \rightarrow e^{-\beta c k}$ on the right side of Eq.~(\ref{eq:kin2}),  
while $\mathcal{U}_0(k,k')$ is symmetrically approximated by its leading behavior 
as  $\beta c k \rightarrow  \infty$,  (see Fig.~\ref{fig:U0quantum}) 
\begin{eqnarray}
\label{eq:quap}
\mathcal{U}_0(k,k') &\sim& \frac{9}{64 \pi^2 m n}\left( e^{-\beta c k} (k - k')^2 \theta(k-k') \right.  \nonumber \\ 
&&+  \left.
 e^{-\beta c k'} (k - k')^2 \theta(k'-k) \right), 
\quad \beta c k \rightarrow \infty .
\end{eqnarray}
The relative size of the second term is $O(e^{-\beta c (k'-k)})$. 
The corresponding subleading contribution to the collision term may be evaluated by 
using the Watson's lemma~\cite{Bender}
\begin{eqnarray}
&& \frac{9}{16 \pi m n} \int_k^\infty e^{-\beta c k'}(k-k')^2 k'^3 \left(
 \mathcal{A}(k',t)  - \mathcal{A}(k,t)  \right)  dk' \sim  \nonumber \\ 
 && \qquad  \qquad
\frac{27 e^{-\beta c k} k^3}{8 \pi m n \beta^4 c^4}  
\partial_k \mathcal{A}(k,t) ,  \quad \beta \to \infty. 
\end{eqnarray}
Although the approximation that arises by ignoring this contribution is necessarily 
non-energy conserving, it may be interesting to explore its consequences,  
since we would expect this drawback to have a negligible effect as $T \to 0$. 
Therefore we consider the evolution equation 
\begin{eqnarray}
\label{eq:T0}
 \partial_t \mathcal{A}(k,t) &=& \frac{9}{16 \pi m n}\frac{1}{k} \int_0^k (k-k')^2 k'^3 \left(
 \mathcal{A}(k',t)  - \mathcal{A}(k,t)  \right)  dk'  \nonumber \\ 
  &=& 
  \frac{9 k^5}{16 \pi m n}   \int_0^k \left(\frac{k'}{k}\right)^6 \left( \frac{k}{k'} -1\right)^2 
  \mathcal{A}(k',t)   \frac{dk'}{k'}   \nonumber \\ 
  &&-\frac{3 k^5}{320 \pi m n} \mathcal{A}(k,t) , 
\end{eqnarray}
where, for convenience,  we have written the gain term as a convolution. 
Unsurprisingly we see that the loss term reduces to 
$-\Gamma_{\mathrm{B}}(k) \mathcal{A}(k,t)$ where 
$\Gamma_{\mathrm{B}}(k)  =  3 k^5/(320 \pi m n)$ 
is precisely the decay width of Beliaev damping at $T=0$~\cite{Beliaev}.  

Although in this regime, $n_0 \sim e^{-\beta c k} \ll 1$,  
we are not in the vicinity of a power law spectrum as 
those appearing in the wave turbulence framework, 
the above linear operator 
is homogeneous  of degree $h=-5$. 
Therefore,  we can  analyze 
the evolution equation with methods similar to the ones we have described before. 
We define the Mellin function corresponding to Eq.~(\ref{eq:T0})
\begin{eqnarray}
W(s) &=& -1 + 60   \int_1^\infty  \frac{(x-1)^2}{x^6} x^{s-1} dx  \nonumber \\ 
         &=& - \frac{s (s^2 - 15 s + 74)}{(s-4)(s-5)(s-6)}, \quad \mathrm{Re}\, s < 4 . 
\end{eqnarray}
The interval of zero winding within  the strip of analyticity is $(-\infty,0)$. 
Rewritten in terms of the Mellin-Laplace image $\mathcal{F}(s,t)$ of $\mathcal{A}(k,t)$ and 
the Mellin image $\Psi(s)$ of the initial condition $\mathcal{A}(k,0)$,  the evolution 
equation reads 
\begin{equation}
\lambda \mathcal{F}(s-5,\lambda) = v_\mathrm{B} W(s) \mathcal{F}(s,\lambda) + \Psi(s - 5) , 
\quad \mathrm{Re}\, s < 0, 
\end{equation}
where $v_\mathrm{B}= 3/(320\pi m n)$. 
Now we seek an appropriate solution  of the difference equation 
$\mathcal{B}(s-5) = -W(s) \mathcal{B}(s)$
in the strip of zero winding,  $\mathrm{Re}\,  s \in (-\infty, 0)$,
and we write the solution $\mathcal{F}(s,\lambda)$ 
in the factorized form~(\ref{eq:conv}). 
To establish this particular solution we require that 
$\mathcal{B}(s)$ has neither zeros nor poles for  $\mathrm{Re}\,  s \in (-\infty, 0)$. 
This requirement is consistent with the location of the zero of the Mellin function at $s=0$. 
It must produce only a simple pole of $\mathcal{B}(s)$  at $s=0$ contained in  
the strip $\mathrm{Re} \in (-\infty, 0]$. The base function turns out to be 
\begin{equation}
\mathcal{B}(s) = \frac{\Gamma\left(-\frac{s}{5} \right)
                                       \Gamma\left(\frac{15}{10}-  i \frac{\sqrt{71}}{10} - \frac{s}{5} \right)
                                       \Gamma\left(\frac{15}{10}+ i \frac{\sqrt{71}}{10} - \frac{s}{5} \right)}
                                       {\Gamma\left(\frac{4-s}{5}\right)\Gamma\left(\frac{5-s}{5}\right)
                                       \Gamma\left(\frac{6-s}{5}\right)}  , 
\end{equation}
and  the inverse images of $\mathcal{B}$ and $1/\mathcal{B}$ are expressed as Mellin-Barnes integrals 
along a vertical path with  $\mathrm{Re}\,  s < 0$. 
The inverse of $\mathcal{B}(s)$ may be computed 
by making the splitting $(\mathcal{B}(s) -1) + 1$ which 
produces  a Meijer $G$ function and a delta function: 
\begin{equation}
\mathcal{Z}(k)  = 5\, G^{0,3}_{3,3}
\left[ k^{5} \left\vert
{1, \frac{-5 + i \sqrt{71}}{10}, \frac{-5 - i \sqrt{71}}{10} \atop 0, \frac{1}{5}, -\frac{1}{5} } 
 \right.
\right] + \delta(k-1) = \mathcal{Z}_1(k)+\delta(k-1). 
\end{equation}
It remains to find the second factor $Q(k)$ of the convolution which 
gives the solution of the initial value problem 
\begin{equation}
\label{eq:solqu}
\mathcal{A}(k, t) =  \int_0^\infty \mathcal{Z}\left(\frac{k}{q} \right) \exp\left(-v_\mathrm{B} q^5 t\right)\, Q(q) \frac{dq}{q}. 
\end{equation}
In particular, if the initial perturbation is $\mathcal{A}(k, 0) = C\, \delta(k-k_0)$, the inversion  of 
$\Psi(s)/\mathcal{B}(s)$ yields $Q(k)$ in the form
\begin{equation}
Q(k) =  \frac{5 C}{k_0} G^{0,3}_{3,3}
\left[ \frac{k^5}{k_0^5} \left\vert
{ 0, \frac{1}{5}, -\frac{1}{5} \atop 1, \frac{-5 + i \sqrt{71}}{10}, \frac{-5 - i \sqrt{71}}{10}  } 
 \right.
\right]   +C\, \delta(k-k_0) = Q_1(k) +  C\, \delta(k-k_0). 
\end{equation}

Since all singularities of $\mathcal{B}(s)$ and $1/ \mathcal{B}(s)$ are located in the half-plane 
$\mathrm{Re}\, s \geq 0$, the Meijer function $G^{0,3}_{3,3}(z)$  vanishes if  $0 \leq z < 1$. 
Some explicit limiting forms are
\begin{eqnarray}
\label{eq:G1}
G^{0,3}_{3,3}
\left[ z \left\vert
{ 0, \frac{1}{5}, -\frac{1}{5} \atop 1, \frac{-5 + i \sqrt{71}}{10}, \frac{-5 - i \sqrt{71}}{10}  } 
 \right.
\right]  &\sim&  -M \left(z- 1\right), \quad z \to 1^+,  \\ 
\label{eq:G2}
 G^{0,3}_{3,3}
\left[ z \left\vert
{1, \frac{-5 + i \sqrt{71}}{10}, \frac{-5 - i \sqrt{71}}{10} \atop 0, \frac{1}{5}, -\frac{1}{5} } 
 \right.
\right]  &\sim& \frac{6\sqrt{10-2 \sqrt{5}}}{5 \cosh (\sqrt{71}\pi/10)}, \quad z \to \infty , 
\end{eqnarray}
where $M \approx 0.48$. 
Consequently, the solution may be rewritten as 
\begin{eqnarray}
\label{eq:solqu2}
\mathcal{A}(k, t) &=&  \int_0^k \mathcal{Z}_1\left(\frac{k}{q} \right) \exp\left(-v_\mathrm{B} q^5 t\right)\, Q(q) \frac{dq}{q} + 
      \exp\left(-v_\mathrm{B} k^5 t\right) Q(k) \nonumber \\
      &=& \frac{C}{k_0} \mathcal{Z}_1\left(\frac{k}{k_0} \right) \exp\left(-v_\mathrm{B} k_0^5 t\right) \theta(k-k_0) + 
      \exp\left(-v_\mathrm{B} k^5 t\right) Q(k) \nonumber \\ 
      &&+\int_0^k \mathcal{Z}_1\left(\frac{k}{q} \right) \exp\left(-v_\mathrm{B} q^5 t\right)\, Q_1(q) \frac{dq}{q}  . 
\end{eqnarray}
It is interesting to find the leading behavior of the last integral as $t \to \infty$. 
To apply the Watson's lemma, we use  Eq.~(\ref{eq:G1}) and thus we obtain
\begin{eqnarray}
\int_0^k \mathcal{Z}_1\left(\frac{k}{q} \right) \exp\left(-v_\mathrm{B} q^5 t\right)\, Q_1(q) \frac{dq}{q}  
&\sim& -\frac{5 C M}{k_0} \mathcal{Z}_1\left(\frac{k}{k_0} \right) \theta(k-k_0)  \nonumber \\
&&\times
 \int_{k_0}^\infty \exp\left(-v_\mathrm{B} q^5 t\right) \left(\frac{q^5}{k_0^5}-1 \right) \frac{dq}{q} \nonumber \\ 
 &\sim& -\frac{C M}{k_0} \mathcal{Z}_1\left(\frac{k}{k_0} \right)  
 \frac{\exp\left(-v_\mathrm{B} k_0^5 t\right) }{ v_\mathrm{B}^2 k_0^{10}\, t^2}  \nonumber \\ 
&&\times \theta(k-k_0),  
 \quad t \to \infty. 
 \end{eqnarray}
 We see  that this integral corresponds to a subleading term of $O((\Gamma_\mathrm{B}(k_0) t)^{-2})$ in the 
expression for $\mathcal{A}(k,t)$. 
 Noting that when $k > k_0$ the second term in Eq.~(\ref{eq:solqu2}) proportional to $Q_1(k)$
 is subleading with respect to the first one,  we obtain the final result  
 for the long time asymptotics of the solution
\begin{eqnarray}
\mathcal{A}(k, t) &\sim& \frac{C}{k_0} \mathcal{Z}_1\left(\frac{k}{k_0} \right) 
\exp\left(-v_\mathrm{B} k_0^5 t\right) \theta(k-k_0) \nonumber \\ 
&&+ 
 C \exp\left(-v_\mathrm{B} k_0^5 t\right) \delta(k-k_0), 
\quad t \to \infty, 
\end{eqnarray} 
or 
\begin{equation}
\mathcal{A}(k, t) \sim \frac{6 C\sqrt{10-2 \sqrt{5}}}{k_0 \cosh (\sqrt{71}\pi/10)}
\exp\left(-v_\mathrm{B} k_0^5 t\right) , 
\quad t \to \infty,  \quad k \gg k_0. 
\end{equation}


\section{\label{sec:end} Concluding remarks}

In this paper we have studied the temporal evolution of 
a perturbation of the equilibrium distribution of noncondensate atoms in
Bose gases as described by a kinetic 
equation including only collisions 
between condensate and noncondensate atoms. 
Due to the difficulty to address this problem without any simplification, 
we have considered some approximations to the kinetic equation 
in  different temperature regimes.
The evolution equations that arise by approximating the full kinetic equation 
when the energy of an excitation is small compared to 
the thermal energy turn out to have definite homogeneity properties, 
and may be regarded as kinetic equations for waves 
in the limit of large occupation numbers. 
This occurs near the critical temperature when the dispersion law  is  quadratic, 
and near zero temperature for Bogoliubov excitations of very low energy. 
Notably, by using the explicit expression that determines the degree of homogeneity,  
one may easily show the dominance of $C_{12}$ over the binary collision term $C_{22}$  
at low momentum. This provides a justification to our approach (i. e. dealing only with $C_{12}$).
In the opposite quantum regime where 
the occupation number of noncondensate atoms decreases exponentially with the momentum, 
we recover unexpectedly a kinetic equation of the same type of the 
wave-turbulence equations in the previous situations.  

To obtain the solution of the initial value problem in all cases, 
we have applied the 
method developed by  Balk and Zakharov~\cite{book,Balk}  to 
analyze the behavior  of weak-turbulent media near Kolmogorov spectra. 
This is a scarcely used technique due to the  
difficulties in the analytical calculation of the Mellin function and the base function, mainly for 
binary processes $2 \leftrightarrow 2$ in turbulent media. 
Here, however,  this task is made easier  as we deal with splitting processes $1 \leftrightarrow 2$ and 
the background distribution corresponds to thermal equilibrium. 

Just below the critical temperature when the condensate is small, 
the isotropic perturbations and the time derivative of the condensate density
vanish for long time according to  power laws, 
while anisotropic perturbations are unstable.   
In  the thermal regime at low temperature, the requirement of energy conservation
leads to a modification of the naive form of the evolution equation,  and  indeed an 
additional source term is needed. 
All solutions of the correct equation decay exponentially for large time and 
the time derivative of the condensate density tends  to zero according to a power law. 
In the quantum regime at very low temperature 
the large time behavior of perturbations  for the non-condensed dynamics turns out to be 
exponentially decaying due to  Beliaev damping processes. 
 
Finally, one may contemplate the possibility of using a similar analysis to that 
we have presented in this paper to evaluate estimations of 
the transport coefficients.


\section*{Acknowledgments}

The work of M.E. is supported by the Spanish MICINN under Grant  MTM2008-03541 and by the Basque Government under Grant No. IT-305-07.
The work of F.P. has been partially supported by Project CBDif-Fr ANR-08-BLAN-0333-01.
The work of M.V. is partially  supported by  the Spanish 
Ministry of Science and Technology 
under Grant  FPA2009-10612, 
the Spanish Consolider-Ingenio 2010 Programme CPAN (CSD2007-00042) and 
the Basque Government under Grant No. IT559-10.


\appendix*
\section{Low momentum analysis of the linearized Boltzmann equation}
\label{sec:app}

In this appendix we give some details about the derivation of 
the limiting form of the collision term for small momentum.  
From Eq.~(\ref{eq:UU}) in the regime near the critical temperature, and setting $u = \cos \theta_{\bm{k} \bm{k}'}$,  
the following expression for $\mathcal{U}(\bm{k}, \bm{k}')$ follows easily
\begin{eqnarray}
\frac{1}{16 n_\mathrm{c} a^2 m^{-2}}\mathcal{U}(\bm{k}, \bm{k}') &=& 
\left[ n_0(\omega(k))[1+n_0(\omega(k'))][1+n_0(\omega(k)-\omega(k'))]  \right. \nonumber \\
 && \times \left.
  \frac{m \theta(k-k')}{k k'}\delta\left(u - \frac{k'}{k}\right) + (k \leftrightarrow k')  \right] \nonumber \\ 
 &&- n_0(\omega(k)+\omega(k'))[1+n_0(\omega(k))][1+n_0(\omega(k'))]  \nonumber \\ 
&& \times \frac{m}{k k'}\delta(u) , 
 \end{eqnarray}
where $n_0$ is the Bose distribution function with zero chemical potential
 \begin{equation}
 n_0(\omega) =  \frac{1}{e^{\beta \omega} -1} . 
 \end{equation}
The coefficients $\mathcal{U}_l(k,k')$ are easily evaluated using the standard formula
\begin{equation}
\mathcal{U}_l(k,k') =  \frac{2 l+1}{2} \int_{-1}^1 \mathcal{U}(\bm{k}, \bm{k}') P_l(u)  du .
\end{equation}

To extract the low momentum behavior of $J_0^<(k,q)$ and  $J_0^>(k,q)$ we first perform exactly the 
integrations of Eqs.~(\ref{eq:J1}) and~(\ref{eq:J2}). 
These may be expressed in closed form 
in terms of  polylogarithmic functions $\mathrm{Li}_n(x)$, 
where $x$ is an exponential of the some combination of reduced energies. 
As an example,  the expressions  for $J_0^{<,>}(k,q)$   near the critical temperature read
\begin{eqnarray}
J_0^<(k, q)  &=&   
-\frac{4\pi n_\mathrm{c} a^2}{\beta m^2} 
\frac{e^{\beta \omega(k)}}{(e^{\beta \omega(k)}-1)^2 k}
 \nonumber \\
 &&\times \left[ \beta^2 q^4 - 4 \beta m q^2  
 \ln \frac{1 - e^{-\beta (\omega(k)-\omega(q))}}{1 - e^{-\beta (\omega(k)+\omega(q))}}   \right. \nonumber \\ 
 && - 8 m^2 \mathrm{Li}_2\left( e^{-\beta (\omega(k)-\omega(q))}  \right) 
        - 8 m^2 \mathrm{Li}_2\left(e^{-\beta (\omega(k)+\omega(q))}  \right)   \nonumber \\
         &&\left.
                       + 16 m^2 \mathrm{Li}_2\left(e^{-\beta \omega(k)}  \right)   \right] \theta(k-q) , \\ 
J_0^>(k, q)  &=&   
\frac{8\pi n_\mathrm{c} a^2}{\beta m^2} 
\frac{e^{\beta \omega(k)}}{(e^{\beta \omega(k)}-1)^2 k}
 \nonumber \\
 &&\times \left[ \beta^2 k^4 - 2 \beta m q^2  
 \ln \frac{\left(1 - e^{-\beta (\omega(q)-\omega(k))} \right)\left(1 - e^{-\beta (\omega(k)+\omega(q))} \right)}
 {\left(1 - e^{-\beta \omega(q)}\right)^2}   \right. \nonumber \\ 
 && + 4 m^2 \mathrm{Li}_2\left( e^{-\beta (\omega(q)-\omega(k))} \right) 
        + 4 m^2 \mathrm{Li}_2\left(e^{-\beta (\omega(k)+\omega(q))} \right)   \nonumber \\
         &&\left.
                       - 8 m^2 \mathrm{Li}_2\left(e^{-\beta \omega(q)}  \right)   \right] \theta(q-k) .                        
\end{eqnarray}
The form of other kernels $J_l^>(k,q)$ with $l \neq 0$ are similar. 
If we substitute  the asymptotic expansion of $\mathrm{Li}_n(x)$ near $x=1$, 
\begin{equation}
\mathrm{Li}_2(x) \sim \frac{\pi^2}{6} + \ln (1-x) \sum_{n=1}^\infty \frac{(1-x)^n}{n} 
- \sum_{n=1}^\infty \frac{(1-x)^n}{n^2} , \quad x \to 1,  
\end{equation}
we obtain   
\begin{eqnarray}
J_0^<(k, q)  &\sim& 
\theta(k-q)\frac{64\pi n_\mathrm{c} m a^2}{\beta^2} \frac{1}{k^3}  \ln\left(1 - \frac{q^4}{k^4}  \right) ,  
\quad k, q \rightarrow 0 , \\ 
J_0^>(k, q)  &\sim&
 \theta(q-k)\frac{64\pi n_\mathrm{c} m a^2}{\beta^2} \frac{1}{k^3}  \ln\frac{q^2+k^2}{q^2-k^2} , 
\qquad k, q \rightarrow 0. 
\end{eqnarray}
These results yield the expression of $\mathcal{H}_0(x)$  given in Eq.~(\ref{eq:h0}), 
once the factor $64\pi n_\mathrm{c} m a^2/\beta^2 \times \sqrt{\beta/m} \times \beta/2m$ 
is extracted to define the time scale $\tau$ in Eq.~(\ref{eq:ectau}). 

Near zero temperature the  expression for $\mathcal{U}(\bm{k}, \bm{k}')$ reads
 \begin{eqnarray}
\mathcal{U}(\bm{k}, \bm{k}') &=& 
\left[\frac{9  (k-k')^2 \theta(k-k')}{32\pi^2 m n}  
  n_0(\omega)[1+n_0(\omega')][1+n_0(\omega-\omega')]  \right. \nonumber \\ 
  && \left.  + (k \leftrightarrow k')  \right]  \delta(u - 1 + 0) \nonumber \\ 
 &&-\frac{9  (k+k')^2 }{32\pi^2 m n}  
  n_0(\omega+\omega')[1+n_0(\omega)][1+n_0(\omega')]  \delta(u - 1 + 0). 
 \end{eqnarray}
We note that the $\delta(u - 1 + 0)$ arise because the dispersion law $\omega = c k + \alpha k^3$  
is not strictly linear and, since $\alpha>0$, the angle between the wave vectors is close to zero. 
The above expression leads directly to the limiting forms given  in Eqs.~(\ref{eq:class}) and~(\ref{eq:quap}).



\section*{References}
\begin{thebibliography}{99}

\bibitem{Kirkpatrick1} T.~R.~Kirkpatrick and J.~R.~Dorfman, Phys. Rev. A {\bf 28}, 2576 (1983).

\bibitem{Kirkpatrick2} T.~R.~Kirkpatrick and J.~R.~Dorfman, J. Low Temp. Phys. {\bf 58}, 301 (1985); 
{\bf 58}, 399 (1985);  {\bf 59}, 1 (1985); 

\bibitem{Zaremba} E.~Zaremba, T.~Nikuni and A.~Griffin, J. Low Temp. Phys. {\bf 116}, 277 (1999). 

\bibitem{book2} A.~Griffin, T.~Nikuni, and E.~Zaremba,  {\it Bose-Condensed Gases at Finite Temperatures} 
(Cambridge University Press: Cambridge,  2009).

\bibitem{Imamovic} M.~Imamovic-Tomasovic and A.~Griffin, J. Low Temp. Phys. {\bf 122}, 617 (2001). 

\bibitem{Newell} S.~Dyachenko, A.~C.~Newell, A~Pushkarev, and V.~E.~Zakharov, Physica D {\bf 57}, 96 (1992).

\bibitem{book} V.~E.~Zakharov, V.~S.~L'vov and G.~Falkovich,  {\it Kolmogorov Spectra of Turbulence I} 
(Springer-Verlag, Berlin, 1992).

\bibitem{Eckern} U.~Eckern, J. Low Temp. Phys. {\bf 54}, 333 (1984); 

\bibitem{Hohenberg} P.~C.~Hohenberg and P.~C.~Martin, Ann. Phys. (N.Y.) {\bf 34}, 291 (1965). 

\bibitem{Beliaev} S.~T.~Beliaev, Zh. Eksp. Teor. Fiz. {\bf 34}, 433 (1958) [Sov. Phys. JETP {\bf 7}, 299 (1958)].

\bibitem{Son} A.~H.~Mueller and D.~T.~Son,  Phys. Lett. B {\bf 582}, 279 (2004). 

\bibitem{Balk} A.~M.~Balk and  V.~E.~ Zakharov,  {\it Amer. Math. Soc. Transl. (2)} {\bf 182},   31 (1998).

\bibitem{Valle} M.~Escobedo and M.~A.~Valle, Phys. Rev. E {\bf 79}, 061122 (2009).

\bibitem{Bender} C.~M.~Bender and S.~A.~Orszag, {\it Advanced Mathematical Methods for Scientists and Engineers} 
(McGraw-Hill, New York, 1978).


\end{thebibliography}
\end{document}